# Literature-based knowledge discovery: the state of the art

[1]Xiaoyong Liu, [2]Hui Fu

[1,]Department of Computer Science, Guangdong Polytechnic Normal University, Guangzhou, Guangdong, 510665,China

liugucas@gmail.com

[*2,]Department of Computer Science, Guangdong Polytechnic Normal University, Guangzhou, Guangdong, 510665,China

lxyong420@126.com

## *Abstract*

*Literature-based knowledge discovery method was introduced by Dr. Swanson in 1986. He hypothesized a connection between Raynaud's phenomenon and dietary fish oil, the field of literature-based discovery (LBD) was born from then on. During the subsequent two decades, LBD's research attracts some scientists including information science, computer science, and biomedical science, etc.. It has been a part of knowledge discovery and text mining. This paper summarizes the development of recent years about LBD and presents two parts, methodology research and applied research. Lastly, some problems are pointed as future research directions.*

## 1. Introduction

Literature-based discovery (LBD) is a new method of knowledge discovery and text mining[1][2], which is presented by Don R. Swanson in 1986. During studying medical literatures, Swanson found the relationship between fish oil and raynaud's disease. Some literatures depicted that there is abnormal in blood of some patients with raynaud's disease (A), such as blood viscosity (B) raising. He found that fish oil (C) can reduce blood viscosity in other literatures. Because no literatures proved the relationship between fish oil and raynaud's disease at that time, so he made a hypothesis that fish oil can cure raynaud's disease. After two years, the hypothesis was proved by clinical trials [3]. Swanson designed and developed a software - ArrowSmith to analysis no-related literatures automatically and find the complementary structure of literatures in 1991.

There are mainly two methods in LBD study, namely "open discovery method" and "closed discovery method"[4]. Open discovery method is the stage of hypothesis providing, i.e. from A to C. Normal process of this method is that, firstly using term A to search literatures, namely literatures A, and then identifying terms B in literatures A, thirdly using term B to search literatures again, namely literatures B, fourthly identifying terms C in literatures B, finally build the linkage from A to C by the term B if A and C aren't co-occurrence in the same literatures. Open discovery method can find a new treatment of disease. Open discovery method can be depicted as fig.1.

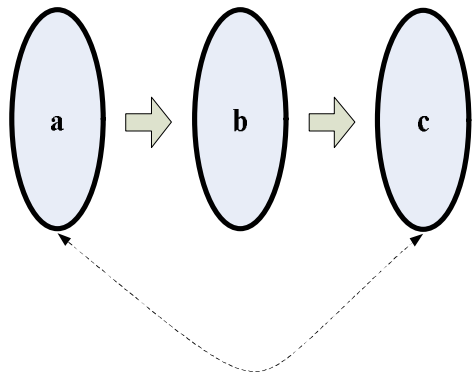

**Fig.1** Open discovery method of LBD

Closed discovery method is the stage of hypothesis verifying. If researchers have get hypothesis, they can verify them by closed method of LBD. They can search literature A and literature C by term A and term C, and find out whether there is same term B in literatures. Closed discovery method can be depicted as fig.2.

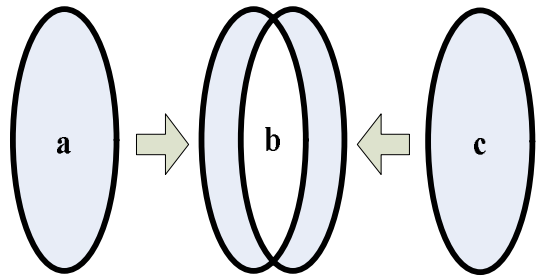

**Fig.2** Closed discovery method of LBD

The ABC knowledge discovery model, which finds out implicit knowledge among no-related literatures, has been of great concern to researchers in information science, computer science and biology science, etc. This paper tracks research progress about LBD study in recent years, and gives a summary from two parts, methodology research and applied research. Lastly, some problems of recent study are indicated and analyzed.

## 2. Methodology research

### 2.1. literature related discovery-LRD

Kostoff provided a novel method of LBD, namely LRD[5], in 2008. Literature-related discovery (LRD) is the linking of two or more literature concepts that have heretofore not been linked (i.e., disjoint), in order to produce novel, interesting, plausible, and intelligible knowledge (i.e.,potential discovery) [6]. Reference [5] introduced the methodology of LRD. His team used five instances to verify the effective of LRD[7][8][9][10][11].

A credible method of LBD needs multiple literatures with four characters [5], as follow,

(1) These multiple literatures should be complementary for solving the problem. If not unique information of each literature, the problem can't be solved.
(2) These multiple literatures should be not been linked (i.e., disjoint).
(3) These multiple literatures should be as comprehensive as possible.
(4) These multiple literatures with unique information must be linked to form a whole that is greater than the sum of its parts.

LRD overcame two roadblocks in classical LBD[6],
(1) Use of numerical filters that are unrelated to generating discovery
(2) Excessive reliance on literatures directly related to the problem literature of interest

There are two improvements in LRD. One strategy is using semantic filters to remove non-related terms, and the other is classifying literatures into core literatures and expanded literatures. Core literatures are those literatures directly related with query. Core literatures are used to obtain technical infrastructure or structure. Expanded literatures are those literatures related with generalizing query. Expanded literatures are used to generate potential discovery candidates.

Generally, The flowchart of LRD is showed in fig.3.

## 2.2. LBD based natural language processing

D. Hristovski and C. Friedman introduced a method of LBD which is using natural language processing (NLP) technology to obtain semantic relationship [12][13][14][15]. They reported on an application of LBD that combines two NLP systems: BioMedLEE and SemRep, which are coupled with an LBD system called BITOLA. The two NLP systems complement each other to increase the types of information utilized by BITOLA. They also discussed issues associated with combining heterogeneous systems. Initial experiments suggest this approach can uncover new associations that were not possible using previous methods. The new paradigm of LBD is showed in fig.4. In 2010, his team developed an improvement system of BITOLA, namely Sempt, to combine semantic relations and DNA microarray data for novel hypotheses generation [16]. However, Kostoff provided that "Although the approach makes use of additional information through the associations, it is still fundamentally a co-occurrence-based concept, with all the deficiencies mentioned previously."[17]

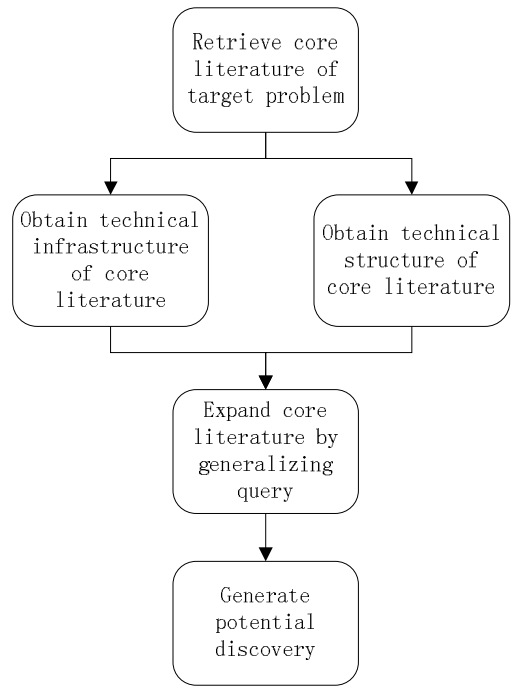

**Fig.3** The flowchart of LRD

## 2.3. Semantic-Based Association Rule: Bio-SARS

Xiao hua Hu presented a new LBD method based on semantic analysis, namely Biomedical Semantic-based Association Rule System (Bio-SARS) [18][19], that significantly reduce spurious, useless and biologically irrelevant connections through semantic filtering. Compared to other approaches such as LSI and traditional association rule-based approach, his approach generated much fewer rules and a lot of these rules represent relevant connections among biological concepts. His method didn't explore semantic relationship between terms, but restricted the semantic type of termsand reduced implicit related knowledge. The semantic types and semantic relationships of UMLS are showed by fig.5.

This paper lists some main LBD system and presents their difference in table 1.

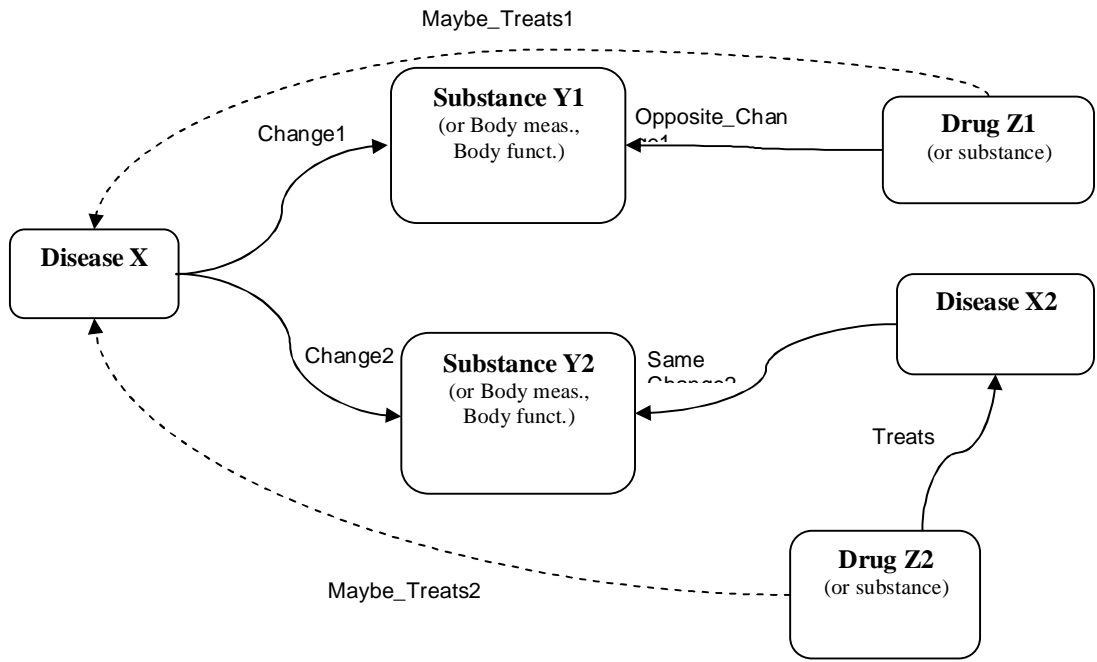

**Fig.4** The paradigm of BITOLA[12]

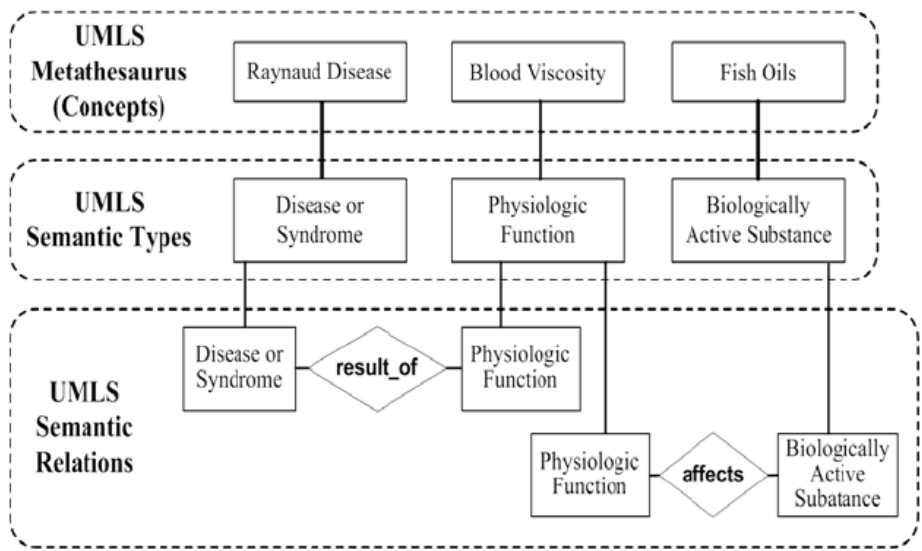

**Fig.5.** Hierarchical structure of UMLS[19]

**Table 1.** Comparation of the LBD system in recent years

| System | Discovery Method | Terms | The method of Concept Capturing | The method of terms set reducing |
|---|---|---|---|---|
| LitLinker[20] | Open | Mesh Terms[I] | Using MetaMap[II] | Removing general terms (such as, medicine, disease, etc.) |
| Literby[21] | Two-stage | Mesh Terms | Using MetaMap | Using semantic type of UMLS[III] to reduce candidate terms. |
| LRD | Open | • Manual input<br>• Query generalizing<br>• Restricting the semantic relationship of terms | • N-Gram model (single, double and three words)<br>• Mesh Terms | • Using semantic type of UMLS to reduce candidate terms<br>• Using semantic type of manual selection to reduce candidate terms |
| RaJoLink[22][23] | Two-stage | Manual input | Mesh Terms | Reducing frequent terms |
| ILP[24] | Closed | Manual input | Using Genia Tagger[IV] | Using improved association rule to reduce candidate term |
| Bio-SARS | Open | Manual input | Mesh Terms | Using semantic type of UMLS and semantic-based association rule to reduce candidate terms. |

## 3. Applied research

LBD is mainly applied in biological medicine or biomedicine. Some researchers used this approach to try to apply in other research area in recent years. All in all, applied research areas have principally as follow,

(1) Biomedicine

Biomedicine is most concentrative an applied area, such as Alzheimer disease and endocannabinoids, migraine and AMPA receptors, schizophrenia and secretin, thalidomide and helicopter pylori, down syndrome and cell polarity, and the cure of autism, etc.. Swanson [25] and Xiaohua Hu [26] used closed discovery method of LBD to identify viruses as potential weapons from complementary literatures.

(2) Water purification

Kostoff firstly applied LRD, a deformation of LBD, to water purification problem [9]. Water purification was the process of removing contaminants from a raw water source. Water purification may remove large particulates such as sand; suspended particles of organic material such as parasites, Giardia, etc.; minerals such as calcium, silica, magnesium; and toxic metals such as lead, copper and chrome. He used LRD to identify purification concepts, technology components and systems that could lead to improved water purification techniques and used expertise to generate potential discovery for water purification.

---

[I] http://www.nlm.nih.gov/mesh/MBrowser.html
[II] http://metamap.nlm.nih.gov/
[III] http://www.nlm.nih.gov/research/umls/
[IV] http://www-tsujii.is.s.u-tokyo.ac.jp/GENIA/tagger/

(3) Social science

M.D. Gordon and N.F. Awad provided that LBD technology could be used to link developed countries and poor developing countries, such as Asia, Latin America and Africa, and to support creative growth of developing countries [27]. It should be possible to allow developing countries to use "leapfrog" technologies that were inconceivable decades ago to support their development. One means of identifying these opportunities is by matching traditional development needs with novel support by connecting previously unrelated literatures. Equally, developed countries could capture unrecognized needs or opportunities that developing countries bring out in growth.

## 4. Problems of recent researches

After summarizing recent study of LBD, it is obviously that there are two problems, which can be as new research directions in future.

### 4.1. Less concern to semantic relationship among terms or concepts

Most studies based on terms co-occurrences to seek out the relationship among terms and make less concern to semantic relationship. Generally, if terms A and terms B co-occurrence in a literature, it doesn't certify that A and B have some kind of semantic relation. They aren't credible knowledge linkages by LBD based on terms co-occurrence.

There are three shortcomings due to less concern to semantic relationship among terms or concepts,

(1) Generate a huge number of possible connections among the millions of biomedical concepts or terms and a lot of these hypothetical connections are spurious, useless and biologically meaningless.

(2) Couldn't reveal the semantic relationship and meanings among implicit terms linkages, i.e. it's impossible that explain the semantic linkage between starting terms A and ending terms C by middle terms B.

(3) Couldn't identify the semantic relationship between terms A and the combination of terms C. In most cases, a kind of substance C is effective for the part of disease A and it doesn't cure completely disease A, while the combination of terms C could be more useful.

### 4.2. Relatively narrow in research areas

Theoretically, the LBD approach could be applied to many domains, but efforts have thus far focused on the biomedicine literatures, specifically MEDLINE or PubMed records. This is not only because MEDLINE records are freely available in electronic format, but also because most LBD efforts identify co-occurring terms as tentative relationships, whether these terms are names or medical subheadings (MeSH). Thus the nature of the association is usually non-specific and is best suited towards associations that are more general in nature [28].

## 5. Conclusions

Knowledge discovery method based on LBD, which is provided by Dr. Swanson, presents a novel approach for information science and computer science. By study of several years, its theory and technology are more and more mature. Some computer software that are represented by Arrowsmith are applied widely to biomedicine science, social science, etc.. That displays the utility value of LBD. However, there are some problems in recent researches of LBD and they will be important research directions in the future.

## 6. Acknowledgement

This work has been supported by grants from Program for Excellent and Creative Young Talents in Universities of Guangdong Province (LYM10097).